\documentclass[sigconf,natbib=true,anonymous=false]{acmart}
\AtBeginDocument{%
  }

\usepackage{amsmath,amsfonts} 
\usepackage{textcomp}
\usepackage{xcolor}
\usepackage{booktabs}
\usepackage{multirow,tabularx}
\usepackage{wrapfig}
\usepackage{epsfig}
\usepackage{makecell}
\usepackage{subfigure}
\usepackage{graphicx}
\usepackage{utfsym}
\usepackage{hyperref}
\usepackage{url}
\usepackage{tablefootnote}
\usepackage{caption}
\usepackage{fancyhdr}
\usepackage{ccicons}

\makeatletter   
\newcommand\figcaption{\def\@captype{figure}\caption}
\newcommand\tabcaption{\def\@captype{table}\caption}
\makeatother

\copyrightyear{2025}
\acmYear{2025}
\setcopyright{cc}
\setcctype{by}
\acmConference[CIKM '25]{Proceedings of the 34th ACM International Conference on Information and Knowledge Management}{November 10--14, 2025}{Seoul, Republic of Korea}
\acmBooktitle{Proceedings of the 34th ACM International Conference on Information and Knowledge Management (CIKM '25), November 10--14, 2025, Seoul, Republic of Korea}\acmDOI{10.1145/3746252.3761429}
\acmISBN{979-8-4007-2040-6/2025/11}

\begin{document}

\title{Exploring the Upper Limits of Text-Based Collaborative Filtering Using Large Language Models: Discoveries and Insights
}

\author{Ruyu Li}
\authornote{Equal Contribution. Authorship order is determined by coin flip.}
\affiliation{
  \institution{Westlake University}
  \city{Hangzhou}
  \country{China}
}
\email{liruyu@westlake.edu.cn}

\author{Wenhao Deng}
\authornotemark[1]
\affiliation{
  \institution{Westlake University}
  \city{Hangzhou}
  \country{China}
}
\email{dengwenhao@westlake.edu.cn}

\author{Yu Cheng}
\affiliation{
  \institution{Westlake University}
  \city{Hangzhou}
  \country{China}}
\email{chengyu@westlake.edu.cn}

\author{Zheng Yuan}
\affiliation{
  \institution{Westlake University}
  \city{Hangzhou}
  \country{China}}
\email{yuanzheng@westlake.edu.cn}

\author{Jiaqi Zhang}
\affiliation{
  \institution{Westlake University}
  \city{Hangzhou}
  \country{China}}
\email{zhangjiaqi@westlake.edu.cn}

\author{Fajie Yuan}
\authornote{Corresponding author. Author contributions: Fajie designed and supervised this research; Ruyu and Wenhao performed this research, in charge of key technical parts; Fajie, Ruyu, and Wenhao wrote the manuscript. Other authors have contributed to some of the experiments in this study.}
\affiliation{
  \institution{Westlake University}
  \city{Hangzhou}
  \country{China}
}
\email{yuanfajie@westlake.edu.cn}

\renewcommand{\shortauthors}{Ruyu Li et al.}

\begin{abstract}

Text-based collaborative filtering (TCF) has emerged as the prominent   technique for text and news recommendation, employing language models (LMs) as text encoders to represent items. However, the current landscape of TCF models mainly relies on the utilization of relatively small or medium-sized LMs. The potential impact of using larger, more powerful language models (such as these with over 100 billion parameters) as item encoders on recommendation performance remains uncertain. Can we anticipate unprecedented results and discover new insights?

To address this question, we undertake a comprehensive series of experiments aimed at exploring the performance limits of the TCF paradigm. Specifically, we progressively augment the scale of item encoders, ranging from one hundred million to one hundred billion parameters, in order to reveal the scaling limits of the TCF paradigm. Moreover, we investigate whether these exceptionally large LMs have the potential to establish a universal item representation for the recommendation task, thereby revolutionizing the traditional ID paradigm, which is considered a significant obstacle to developing transferable “one model fits all” recommender models. Our study not only demonstrates positive results but also uncovers unexpected negative outcomes, illuminating the current state of the TCF paradigm within the community. These findings will evoke deep reflection and inspire further research on text-based recommender systems. Our code, datasets and additional materials are provided at \textcolor{blue}{\url{https://github.com/westlake-repl/TCF}}.
\end{abstract}




\maketitle

\section{Introduction}\label{section:Introduction}
\begin{figure*}[t]
        \centering
        \includegraphics[width=0.92\textwidth]{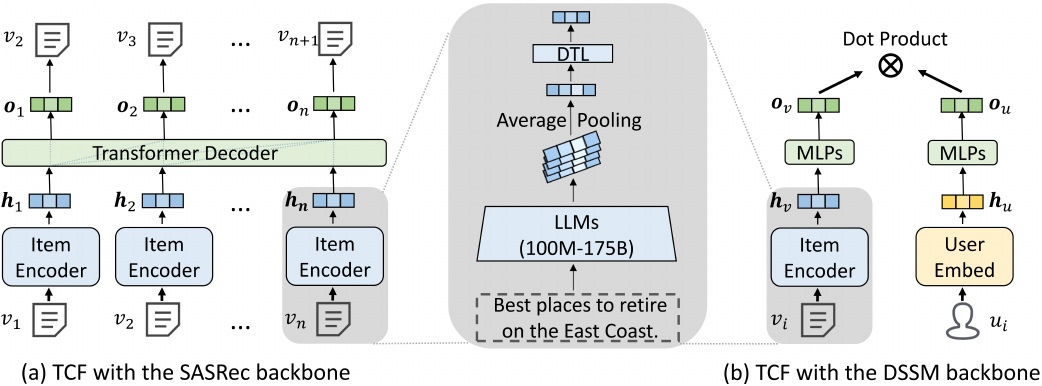} 
        \caption{Utilizing LLMs as item encoders in TCF, with SASRec/DSSM as recommender backbones. The DTL block is the dense dimension transformation layers. Item or text encoder used in this study can be 175B parameters.} 
        \label{Architecture} 
    \end{figure*}

The explosive growth of online text data has emphasized the significance of text content recommendation across various domains, including e-commerce, news recommendation, and social media. Text-based collaborative filtering (TCF) has emerged as a pivotal technology for delivering personalized recommendations to users based on textual data, such as product descriptions, reviews, or news articles~\citep{wu2021empowering,yuan2023go}. The objective of TCF is to accurately capture user preferences and interests from textual data and provide customized recommendations that align with their needs.    TCF typically employs language models (LMs) as item encoders of textual data, integrated into a recommender architecture using collaborative filtering techniques~\citep{rendle2010factorizing,he2017neural,koren2009matrix}  to generate user-item matching scores (see Figure~\ref{Architecture}).
The promising results of TCF have established it as the mainstream approach for text-based recommendation.\footnote{In this paper, we primarily focused on utilizing LMs as the item encoder within the TCF framework. However, it is worth noting that LMs have gained traction as the recommendation (or user) backbone (e.g.,~\citep{li2023gpt4rec}) in recent years, because of the rise of large LMs. Some key results of this alternative paradigm  are presented in our github document (see the link of our abstract), as they are beyond the scope of current study.}

By employing language models  as item encoders, TCF naturally benefits from the latest advancements in natural language processing (NLP). Particularly, in recent years, large LMs (LLMs) such as GPT-3~\citep{brown2020language}, GPT-4~\citep{gpt-4} and LLaMA~\citep{touvron2023llama, touvron2023llama2}  have achieved revolutionary successes in modeling textual data. However, the text encoders utilized in conventional TCF models often consist of relatively small or medium-sized LMs, such as word2vec~\citep{mikolov2013distributed}, BERT~\citep{devlin2018bert}, and RoBERTa~\citep{liu2019roberta}. This limitation may restrict their recommendation capabilities, leading to essential questions: Can TCF achieve exceptional results by leveraging extremely large LMs with tens or hundreds of billions of parameters as text encoders? Is there an upper limit to TCF's performance when pushing the size of the text encoder to its extreme?  Can TCF with the LLMs revolutionize the prevailing ID paradigm and usher in a transformative era akin to the universal foundation models~\citep{bommasani2021opportunities}  in NLP?

Undoubtedly, the above questions play a crucial role in guiding research within the mainstream TCF paradigm.  However, despite numerous TCF algorithms proposed in literature~\citep{wu2021empowering,zhang2021unbert, li2022miner,bi2022mtrec,xiao2022training}, none of them  have explicitly discussed the above questions. Therefore, instead of introducing yet another algorithm, we aim to decipher the classic TCF models via a series of \textit{audacious experiments}  that require immense computational resources. Specifically, we explore the below novel questions.

\textbf{Q1: How does the recommender system's performance respond to the continuous increase in the item encoder's size?  Is the performance limits attainable at the scale of hundreds of billions?} 
To answer it, we perform an empirical study where we systematically increase the size of the text encoder from 100 million (100M for short) to 175 billion (175B).     This study is conducted on three recommendation datasets, utilizing two most representative recommendation architectures: the two-tower  DSSM
~\citep{huang2013learning, huang2020embedding} model and the sequential model SASRec~\citep{kang2018self} with the Transformer~\citep{vaswani2017attention} decoder as the backbone.  

\textit{Novelty clarification}: While the scaling effect has been established in the NLP field, it is important to note that recommender models not only involve the item encoder but also the user encoder. As a result, the potential improvement solely from scaling the item encoder remains unknown.  A  concurrent
preprint~\citep{kang2023llms} by Google teams investigated the impact of scaling the item encoder on explicit rating prediction. However, to  our best knowledge, we are the first to explore the scaling effect in the context of top-$N$ item recommendation  from implicit feedback ~\citep{cremonesi2010performance}.

\textbf{Q2: Can LLMs with over 100 billion parameters generate universal item representations for recommendation?
}
Developing universal foundation models is an ambitious goal of NLP, as previous studies have showed the generality of the representations learned by LLMs across various NLP tasks. However, recommender systems (RS) differ from these objective NLP tasks as they are personalized and subjective. This raises the question whether the LLMs pre-trained on non-recommendation data can produce a universal item representation in the recommendation context.

\textbf{Q3: Can recommender models with a 175B parameter LLM as the item encoder easily beat the simplest ID embedding based models (IDCF), especially  for warm item recommendation?  }
IDCF is a prevailing and the state-of-the-art recommendation paradigm that has dominated the recommender system (RS) community for over a decade, particularly in the non-cold start setting. It produces high-quality recommendations without relying on any item content information. However, recent studies~\citep{ding2021zero,hou2022towards,wang2022transrec,yuan2023go,zhang2024ninerec,cheng2024image} indicate  that ID features are the key barrier to achieving  transferable ``one model fits all'' recommender models (see Figure~\ref{fig:foundation}).  This is because IDs, e.g., userID and itemID, are typically not shareable across different practical platforms.

\textit{Novelty clarification}: While numerous papers have claimed that their proposed TCF  has achieved state-of-the-art performance, it is important to note that many claims are primarily focused on either cold-start scenarios~\citep{zhang2021unbert} or sub-optimal embedding sizes (e.g., fixed small sizes).
However, in order to entirely subvert the ID paradigm, it is crucial to surpass its performance in both cold-start and warm scenarios under the fair evaluation setting. This presents a considerable challenge, which also explains why  few large-scale industrial recommender systems  claimed to completely abandon the itemID features (userID can be represented by itemID sequence) in the non-cold start recommendation setting.

\textbf{Q4: How close is the TCF paradigm to a universal ``one model fits all'' recommender model? }
In addition to its performance benefits, TCF is often lauded for its potential transferability~\citep{zhang2024ninerec,wang2022transrec,fu2024exploring,hou2022towards}, allowing for cross-domain and cross-platform recommendations without relying on shared data. This advantage contributes to the establishment of a universal foundation model ~\citep{bommasani2021opportunities} in the field of recommender systems. Therefore, we aim to study whether TCF, utilizing LLM as the item encoder, exhibits effective transferability, particularly its zero-shot recommendation capability~\citep{ding2021zero}, a prominent research direction in recent years.

If both Q3 and Q4 hold true, LLM will undoubtedly possess the potential to revolutionize the existing recommendation paradigm.\footnote{In this study, our focus is solely on textual items. However, it is crucial to acknowledge that image, audio, and video items can also be represented as texts by leveraging multimodal techniques, e.g.,~\citep{zhu2023minigpt}. That is, TCF paradigm also applies for these scenarios.}  In the future, it is conceivable that similar recommendation scenarios could be addressed with a single recommender model, significantly minimizing the need for redundant engineering efforts. However, so far, whether the RS field can develop universal models similar to the NLP community still remains unknown, and the entire community is unable to give a definitive answer.   
Our primary contribution in this paper is to conduct preliminary research and establish a substantial factual foundation for addressing this question more comprehensively in the near future.

    \section{Background}
        
\textbf{LMs for Text.}  
In recent years, significant progress in LM development has had a profound impact on the field of NLP.
word2vec, developed in 2013, revolutionized NLP by providing a scalable and efficient way of learning word embeddings. Since then, various improvements have been made to word representation models, such as GloVe~\citep{pennington2014glove}, TextCNN~\citep{Kim2015convolutional}, ELMo~\citep{peters2018deep}, etc.  In 2018, the BERT model showed  state-of-the-art performance on a range of NLP tasks  by introducing a pre-training approach based on masked language modeling. BERT and its variants (RoBERTa~\citep{liu2019roberta}, ALBERT~\citep{lan2019albert}, XLNet~\citep{yang2019xlnet},  T5~\citep{raffel2020exploring}, etc.) have become a dominant paradigm in the NLP community in recent years. More recently, ChatGPT\footnote{https://chat.openai.com/}, a conversational LLM model  has gained significant attention for its remarkable performance  in a wide range of language tasks. Along this line, several other notable works have contributed to the advancement of LMs, including the Transformer architecture, the GPT ~\citep{radford2018improving,radford2019language,brown2020language} and
Llama~\citep{touvron2023llama,touvron2023llama2}  models.

\textbf{LMs for Recommender Systems.}
Over the past years, LMs have been widely used in item recommendation tasks, with two main lines of research in this area. The first involves using LMs to represent textual items~\citep{wu2021empowering,wu2019neural,zhang2021unbert,yuan2023go,wu-etal-2020-mind}, while the second involves using LMs as user encoders or recommendation backbones, such as SASRec, BERT4Rec~\citep{sun2019bert4rec}, GRU4Rec~\citep{hidasi2015session}, NextItNet~\citep{yuan2019simple}, P5~\citep{geng2023recommendation} and  GPT4Rec~\citep{li2023gpt4rec}. In this paper, we focus primarily on the first line of research. Among the various item encoders, lightweight word2vec and medium-sized BERT  are the two most popular options. The literature on this topic can further be classified into two categories: applying pre-extracted textual features (equivalent to a frozen text encoder)~\citep{ding2021zero,bi2022mtrec} and end-to-end (E2E) training of text encoders~\citep{yuan2023go,yang2022gram,li2023text}. While E2E training typically achieves better results than using a frozen text encoder, the latter approach is much more computationally efficient than E2E training~\citep{yuan2023go}. 

The success of ChatGPT has  sparked the   use of prompt techniques for personalized recommendations~\citep{gao2023chat,liu2023chatgpt,dai2023uncovering}. This approach can directly utilize the ChatGPT API, eliminating the need for separate model training. It is noteworthy that in recent several months, there has been a significant amount of literature on LLM-based recommender systems, covering a variety of paradigms~\citep{wu2023survey,qin2023large,hou2023large,fan2023recommender,salemi2023lamp}. \textbf{However, this paper specifically concentrates on the utilization of LLM as the item encoder,  a prominent paradigm in the field.}

\section{Preliminary}
\label{Preliminary}

We introduce  some basic notations and describe two typical recommender paradigms: IDCF \& TCF.

\textit{Definition.} We define the set of users as \(U = \{u_1, u_2, ..., u_m\}\) and  the set of items as \(V = \{v_1, v_2, ..., v_n\}\). The user-item interactions are represented by a binary matrix  \(R = \{r_{uv}\}\), where \(r_{uv} \in \{0, 1\}\) indicates whether user \(u\) has interacted with item \(v\).

In the standard collaborative filtering (CF) setup, we represent each user by a vector $\theta_u \in \mathbb{R}^k$ and each item by a vector $\beta_v \in \mathbb{R}^k$. The predicted interaction score between user $u$ and item $v$ is computed as $\hat{r}_{uv} = \theta_u^T \beta_v$. To obtain the user and item vectors, we typically optimize a loss function $l(r_{uv}, \theta_u^T \beta_v)$, where $l$ can either be a pairwise  BPR~\citep{rendle2012bpr} loss or a cross-entropy  loss.

In the popular ID-based CF (IDCF) models,  $\theta_u$ and $\beta_v$, also known as userID and itemID embeddings, can be learned by backpropagating from the user-item interaction data.
Following this path, various recommender models have been developed. For instance, if we use a deep neural network to output the user vector $\theta_u$ and the item vector $\beta_v$, denoted by $g(u_i)$ and $h(v_i)$ respectively, the scoring function becomes $\hat{r}_{uv} = g(u_i) \cdot h(v_i)$, which is known as the two-tower DSSM model. Alternatively, if we represent a user by a sequence of $k$ items that she has interacted with, the scoring function is $\hat{r}_{uv} = G(v_1, v_2, ..., v_k)^T \beta_{v}$, where $G(\cdot)$ is a sequential network, such as SASRec and BERT4Rec.

By utilizing a text encoder $f(v_i)$ to output item representation vectors from the description text, instead of relying on itemID embedding features, the IDCF model 
can be converted into the TCF model, as depicted in Figure~\ref{Architecture}.  Clearly, the only difference between TCF and the typical IDCF model is in the item representation part. In contrast to IDCF, TCF has the advantage of being able to utilize both item textual content features and user-item interaction feedback data. In theory, the text encoder $f(v_i)$ can take the form of any language model, such as a shallow-layer word2vec model, a medium-sized BERT model, or a super-large GPT-3 model. The text encoder $f(v_i)$ can be either frozen or trained jointly with the whole recommender model in an end-to-end (E2E) fashion. 

However, due to computational constraints, most real-world recommender systems adopt a two-stage approach.  In this approach, offline features are extracted in advance from a frozen LM encoder and then incorporated as fixed features into the recommender model during both training and inference stages. This is primarily due to the resource-intensive nature of joint or E2E training of text encoders, which requires substantial computing power and time~\citep{yuan2023go}.
 
\section{Experimental Setups}
\subsection{Datasets, Models and Evaluation}
\begin{table*}
      \caption{Dataset characteristics. Bili8M is mainly used for pre-training to answer Q4.}
      \label{Sample-dataset}
      \centering
      \begin{tabularx}{\linewidth}{p{2.5cm}<{\centering} p{2.0cm}<{\centering} p{2.0cm}<{\centering} p{2.0cm}<{\centering} r l}
        \toprule
        Dataset &\#User   &\#Item  &\#Interaction  & & Item Example \\
        \midrule
        MIND    &200,000  &54,246  &2,920,730  & &{Cincinnati Football History} (News Title) \\
        HM      &200,000  &85,019  &3,160,543  & &{Solid. White. Ladieswear.} (Product Description) \\
        Bili    &50,000   &22,377  &723,071    & &{The last words of The Humans} (Video Title)\\
         Bili8M  &8,880,000  &408,000  &104,450,865 & &{The owl is wearing a skirt} (Video Title)\\
        \bottomrule
      \end{tabularx}
    \end{table*}
\textbf{Datasets.} We evaluate TCF with LLM as item encoders on three real-world text datasets: the MIND news clicking dataset~\citep{wu-etal-2020-mind}, the HM clothing purchase dataset\footnote{\url{https://www.kaggle.com/competitions/h-and-m-personalized-fashion-recommendations}}, and the Bili\footnote{URL: https://www.bilibili.com/. To create this dataset, we randomly crawled short video URLs (with durations of less than 10 minutes) from 23 vertical channels (including technology, cartoons, games, movies, food, fashion, etc.) in Bili. We then extracted the public comments on these videos as positive interactions. Finally, we chronologically combined all user interactions and removed duplicate interactions as the final dataset. This full dataset has been released by ~\citep{zhang2024ninerec}.} comment dataset  from an online video recommendation platform. For MIND, we represent items using their news article titles, whereas for HM and Bili, we utilize the respective title descriptions of clothes or videos to represent the items. Across all datasets, each positive user-item interaction is either a click, purchase, or comment, which serves as an implicit indicator of user preference.


Due to memory issues when comparing to E2E training, we constructed interaction sequences for each user by selecting their latest 23 items.\footnote{The number 23 was selected at random, as we set the maximum sequence length to 20. Specifically, the 21st item was used for prediction within the training set (i.e., $1,2,...,20 \Rightarrow 2,3,...,21$ for an autoregressive model such as SASRec), the 22nd item was assigned to the validation set, and the 23rd item was assigned to the testing set.} 
We exclude users with fewer than 5 interactions as we do not consider cold user settings. Following the basic pre-processing steps, we randomly selected 200,000 users (along with their interactions) from both the MIND and HM datasets, as well as 50,000 users from Bili for our main experiments. Additionally, we have also built a large-scale Bili8M (covering the entire Bili dataset) dataset for pre-training purposes to answer Q4. Datasets statistics are reported in Table~\ref{Sample-dataset}. We report details of them in \textbf{Appendix C}.

\textbf{Models and Training.} To support our main arguments, we selected two most representative recommendation architectures for evaluation: the classical two-tower DSSM model and the Transformer-based  sequential model SASRec. Many new recommender models can be categorized under the DSSM and SASRec frameworks. For instance, numerous advanced CTR  (click-through rate)  prediction or top-$N$ ranking models, despite being single-tower models, are expected to yield similar conclusions as the two-tower DSSM model. Therefore, it will not affect our key conclusions. Likewise, SASRec with a vanilla Transformer backbone can represent multiple multi-head self-attention variants and is still considered state-of-the-art after being developed for over 6 years, see~\citep{narang2021transformer}.

Regarding training, we utilize the popular batch softmax loss~\citep{yi2019sampling}, which is widely adopted in large-scale industrial systems. 
For text encoders, we evaluated nine different sizes of LLM models, ranging from 125M to 175B parameters. 
These LLM models were implemented by Meta AI with the aim of reproducing GPT-3, also known as OPT~\citep{zhang2022opt}. Kindly note that at the time of conducting these experiments, OPT was the only and largest open-source LLM with over 100 billion parameters. However, considering the recent release of more open-source LLMs, we have included additional experiments in our benchmark section (i.e., section~\ref{section-benchmark}), involving these newly available models.
As for hyper-parameters, we first perform a grid search for standard IDCF as a reference,
After determining the optimal hyper-parameters for IDCF, we search them for TCF 
around these optimal values. We report details of them in \textbf{Appendix A}.

\textbf{Evaluation.} We present the performance of all models using the widely adopted top-N ranking metric, HR@10 (Hit Ratio). Due to space constraints and the high consistency observed between HR@10 and NDCG@10 (Normalized Discounted Cumulative Gain), we report all results for NDCG@10 on our \href{https://github.com/westlake-repl/TCF}{\color{blue}{github}} page.
The latest user-item interaction was used for evaluation, while the second-to-last interaction was used for  hyper-parameter searching, and all other interactions were used for training. All items in the pool are used for evaluation, suggested by~\citep{krichene2020sampled}. 

     \begin{figure*}[t]
        \centering
        \includegraphics[width=1\textwidth]{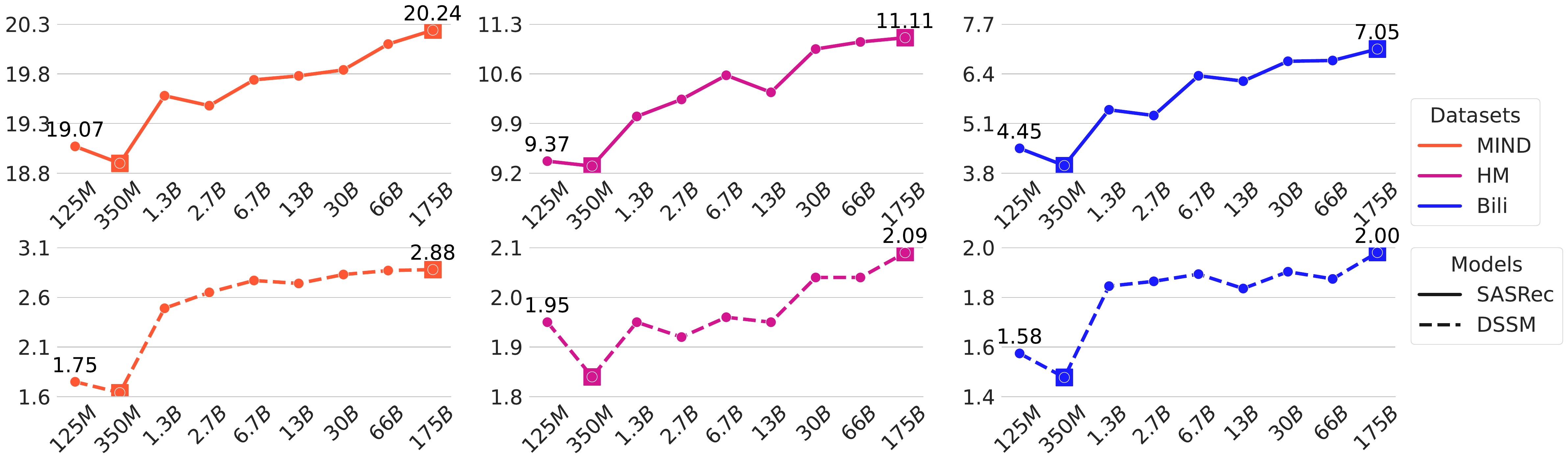} 
        \caption{TCF’s performance (y-axis: HR@10(\%)) with 9 OPT-based text encoders of increasing size (x-axis). SASRec (upper three subfigures) and DSSM (bottom three subfigures) are used as the backbone.} 
        \label{Scaling up} 
    \end{figure*}

\section{Q1: Has the TCF paradigm reached its performance ceiling by scaling item encoders to over 100 billion parameters?}
\label{upperlimit}

To answer Q1, we conduct experiments by increasing the size of text encoders in the TCF models, ranging from 125M to 175B parameters. We use SASRec and DSSM as recommender backbones. 
The results are given in Figure~\ref{Scaling up}. All LMs are frozen for this study, unless explicitly stated otherwise.

As shown,  TCF models generally improve their performance by increasing the size of their text encoders.
For instance, with the SASRec as the backbone, TCF improved the recommendation accuracy from 19.07 to 20.24 on MIND, from 9.37 to 11.11 on HM, and from 4.45 to 7.05 on Bili, resulting in improvements of 6.1\%, 18.6\%, and 58.4\%, respectively. Similar trend can also be made for the DSSM backbone. Additionally, we have also confirmed consistent accuracy improvement from 7B to 70B  by applying LLama2 as the text encoder in Table~\ref{tab:LMcompare}. Meanwhile, the results show that LLama2 with 70B parameters is unable to outperform OPT175B.\footnote{Note that we cannot solely attribute the superior performance of OPT175B over LLama2 to its larger parameter size. The disparity in their training data and training methods also plays a significant role in determining the recommendation accuracy.} 

Based on the observed performance trend, we can conclude that the TCF models' performance has not yet converged when increasing the size of their text encoders, such as from 13B to 175B.  These results suggest that \textbf{(answer to Q1) the TCF model with a 175B parameter LM may not have reached its performance ceiling}. In other words, if we had an even larger LM as the text encoder, TCF's performance could potentially be further improved. \textbf{To the best of our knowledge, this paper provides the first evidence of the scaling effects of LLMs with hundreds of billions of parameters as item encoders for the top-N item recommendation task.}


Interestingly, we find that the TCF model with the 350M parameter LM  exhibits the poorest performance across all three datasets, regardless of whether it uses the DSSM or SASRec backbone. However, the 350M LM is not the smallest text encoder.  This could happen because the scaling relationship between text encoder size and performance is not necessarily strictly linear. However, by examining the pre-training code and official documentation, we discover that the 350M-parameter OPT was  implemented  with several differences compared to all other versions.\footnote{For instance, in all other pre-trained models, the layernorm layer is implemented before the attention layer, while in the 350M model, it is  opposite. Plus, its embedding \&  hidden layer dimensions  are also set differently.} This provides an explanation for our results. Additionally, beyond the discussion scope of this paper,  we also note that TCF utilizing the SASRec backbone shows significantly superior performance compared to TCF with the DSSM backbone. Similar findings were reported in much previous literature~\citep{kang2018self,sun2019bert4rec}. One possible reason for this is that representing users using their interacted items is more effective than using solely the userID feature. Another reason could be that the SASRec architecture, based on the sequence-to-sequence (seq2seq) training approach, is more powerful than the two-tower DSSM architecture.

\section{Q2: Can the 175B LLM achieve universal item representation for RS?}

We are curious about whether a LLM with 175B parameters possess a degree of universality in text encoding. Unlike the objective NLP tasks, here we examine this property using personalized  recommendation as a downstream task. 

Assuming that a $k$-dimensional text representation $\beta_v$ encoded by the 175B parameter LLM is an ideal universal representation, any application involving text representation can directly choose a subset or the entire set of features from $\beta_v$ by providing a weight vector $w$ that represents the importance of these elements, i.e., $y = w^T \beta_v$.
For example, in a basic matrix factorization setting, $w$ could represent user preference weight to item features, i.e. $w = \theta_u$. If all factors of user preference can be observed by the features in  $\beta_v$, we only need to learn their linear relationship. Moreover, for a perfect universal vector  $\beta_v$, using a frozen representation  should be just as effective as fine-tuning it on a new dataset, or even superior to fine-tuning. 

Based on the analysis, we can simply compare the frozen item representation with the fine-tuned item representation to verify our question. 
Note that previous studies such as ~\citep{yuan2023go} have investigated this issue, but they only examined text encoders with a size of 100M parameters.
Given the significantly enhanced representation capabilities of the 175B LM (as shown in Table~\ref{tab:LMcompare}), it is uncertain whether the findings remain consistent when the encoder is scaled up by a factor of 1000.

\begin{figure*}[t]
    \centering
    \includegraphics[width=1.0\textwidth]{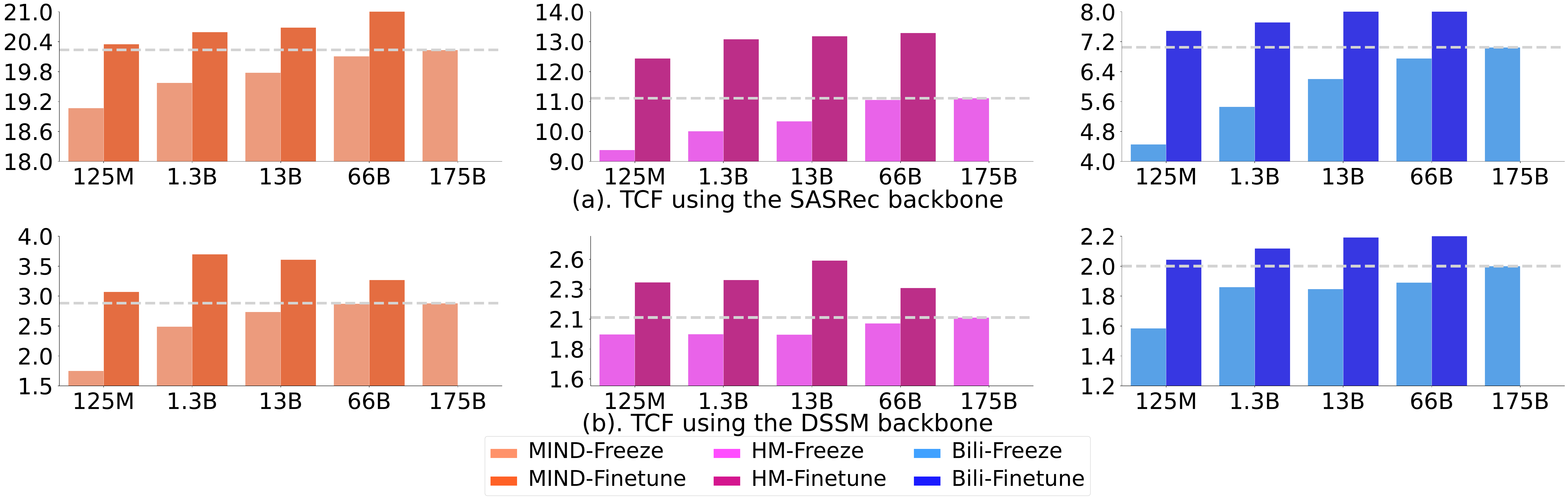} 
    \caption{TCF with retrained LM vs frozen LM (y-axis: HR@10(\%)), where only the top two layers are retrained. The 175B LM is not retrained due to its ultra-high computational cost.} 
    \label{fig:universalrep} 
\end{figure*}

As shown in Figure~\ref{fig:universalrep}, TCF models (both SASRec and DSSM) outperform their frozen versions when the text encoders are retrained on the recommendation dataset. Surprisingly, TCF with a fine-tuned 125M LM is even more powerful than the same model with a frozen 175B LM. 
This result potentially suggests that \textbf{(answer to Q2) even the item representation  learned by an extremely large LM may not result in a universal representation, at least not  for the text recommendation task.} 
Another key insight is that although LLMs have revolutionized so many NLP problems, there is still a significant domain gap between RS and NLP - specifically, inferring user preferences  appears to be more challenging.
We suspect that  the text representation even extracted from the strongest and largest LM developed in the future may not perfectly adapt to the RS dataset. Retraining the LLM on the target recommendation data is necessary for optimal results. 
However, from a positive perspective, since LLMs have not yet reached the performance limit, if future larger and more powerful LLMs are developed, the performance of frozen text representation may become more close to fine-tuning.
For instance, we observe that SASRec with a 175B LM (compared to the 125M LM) is already very close in performance to the fine-tuned 66B LM,  with relative accuracy gaps of 3.92\%, 16\%, 13.5\% on HM, and Bili, respectively.
This is a promising discovery since fine-tuning such a large LM is very challenging in practical scenarios.
Note while we did not fine-tune all layers of the 175B LM, we did assess the performance using medium-sized LMs (including 1.3B, 13B and 30B) by optimizing all layers and the top two layers, which yielded comparable results.

It is worth noting that the above conclusions are based on the assumption that user-item interaction feedback serves as the gold standard for the recommendation task, but this may not always be the case in practice. As a limitation, this study does not address this issue, as the entire theory of modern recommender systems is currently based on this assumption.

\begin{table}[t]
\centering
\caption{Warm item recommendation (HR@10). 20 means items < 20 interactions are removed. TCF\textsubscript{175B} uses the pre-extracted features from the 175B LM. Only the SASRec backbone is reported. }
\label{tab:PLM_warm}
\begin{tabular}{p{0.9cm}<{\centering} p{0.5cm}<{\centering} p{0.5cm}<{\centering} p{0.5cm}<{\centering} p{0.4cm}<{\centering} p{0.4cm}<{\centering} p{0.5cm}<{\centering} p{0.4cm}<{\centering} p{0.4cm}<{\centering} p{0.4cm}<{\centering}}
\toprule
\textbf{Data} &\multicolumn{3}{c}{MIND} &\multicolumn{3}{c}{HM} &\multicolumn{3}{c}{Bili} \\
\cmidrule(lr){2-4} \cmidrule(lr){5-7} \cmidrule(lr){8-10} 
\textbf{\#Inter.} &20 &50 &200 &20 &50 &200 &20 &50 &200 \\
\midrule
IDCF                      &20.56 &20.87 &23.04 &13.02 &14.38 &18.07 &7.89 &9.03 &15.58 \\
TCF\textsubscript{175B}   &20.59 &21.20 &22.85 &12.03 &12.68 &16.06 &7.76 &8.96 &15.47 \\

\bottomrule
\end{tabular}
\end{table}

\section{Q3: Can IDCF be easily surpassed by TCF with a 175B LLM?}
TCF is a classical paradigm for text-based recommendation, while IDCF is the dominant paradigm in the entire field of RS. Can TCF models with a 175B parameter LLM easily beat IDCF models with learnable item embedding vectors? While many prior studies have reported that their TCF models achieved state-of-the-art results, few have explicitly and fairly compared their models with corresponding IDCF counterparts \textit{under the same backbone networks and experimental settings (including samplers and loss functions)}. Moreover, many of them focus on cold item setting, with fewer studies explicitly examining regular (with both cold and warm items) or warm item settings. 
Recently,~\citep{yuan2023go} discovered  that TCF can be comparable to IDCF by jointly training a 100M parameter LM, but frozen LM still significantly underperformed. Therefore, a natural question is whether our conclusions would differ if we use a 1000x larger LLM as the item encoder?

As shown in Table~\ref{tab:IDvsLLM}, we observe that even with the 175B parameter LLM and fine-tuned 66B parameter LLM, TCF is still substantially inferior to IDCF when using DSSM as the backbone. These results are consistent with~\citep{yuan2023go}. As explained, the DSSM architecture and training approach exhibit limited effectiveness in training TCF models. Both the IDCF and TCF models with DSSM perform worse than the seq2seq-based SASRec model. However, \textbf{a notable finding different from \citep{yuan2023go} is that we reveal that TCF with the SASRec backbone performs comparably to IDCF on the MIND and Bili datasets, even when the LLM encoder is \textit{frozen}}, as shown in Table~\ref{tab:IDvsLLM} and~\ref{tab:PLM_warm}. This represents a significant advancement since no previous study has \textit{explicitly} claimed that TCF, by only freezing an NLP encoder (or utilizing pre-extracted fixed representations), can achieve on par performance to its IDCF counterparts specifically in the context of warm item recommendation.\footnote{We simply omit the results for cold item recommendation, as TCF has been consistently showed to outperform IDCF in such settings in numerous literature, e.g., in~\citep{yuan2023go,hou2022towards}.} This is probably because smaller LM-based item encoders in prior literature, such as BERT and word2vec, are inadequate in generating effective text representations comparable to IDCF, see Table~\ref{tab:LMcompare}.

\begin{table}[t]
\centering
\caption{Accuracy comparison (HR@10) of IDCF and TCF using the DSSM \& SASRec backbones. \textit{FR} is TCF using frozen LM, while \textit{FT}  is TCF using fine-tuned LM.}
\label{tab:IDvsLLM}
\begin{tabular}{p{1.1cm}<{\centering} p{0.8cm}<{\centering} p{0.8cm}<{\centering} p{0.8cm}<{\centering} p{0.8cm}<{\centering} p{0.8cm}<{\centering} p{0.8cm}<{\centering}}
\toprule
\multirow{2}{*}{\textbf{Data}} &\multicolumn{3}{c}{SASRec} &\multicolumn{3}{c}{DSSM} \\
\cmidrule(lr){2-4} \cmidrule(lr){5-7}
&IDCF &{{175B}\textsuperscript{\textit{FR}}} &{{66B}\textsuperscript{\textit{FT}}} &IDCF &{{175B}\textsuperscript{\textit{FR}}} &{{66B}\textsuperscript{\textit{FT}}}\\

\midrule
MIND & 20.05 & 20.24 & \textbf{21.07} & \textbf{3.99} & 2.88 & 3.27 \\
HM   & 12.02 & 11.11 & \textbf{13.29} & \textbf{6.79} & 2.09 & 2.35 \\
Bili & 7.01 & 7.05   & \textbf{8.15}  & \textbf{2.27} & 2.00 & 2.04 \\
\bottomrule
\end{tabular}
\end{table}

\begin{figure}[t]
    \centering
    \includegraphics[width=0.47\textwidth]{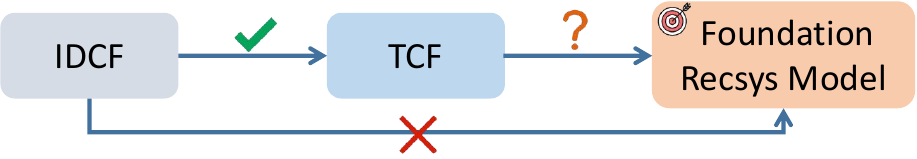} 
    \caption{Route to foundation recommender models (FRM). The cross indicates that the IDCF paradigm have no chance to achieve FRM, the tick indicates that for text-centric RS, TCF can basically replace IDCF, and the question mark indicates that whether the TCF paradigm can achieve the widely recognized FRM remains still unknown for the RS community. } 
    \label{fig:foundation} 
\end{figure}

The reason for the weaker performance of TCF on HM is that textual information alone is insufficient to fully represent the product item, as factors such as price and quality are also critical in enticing user clicks and purchases on HM. However, in the case of news recommendation, we can generally assume that users are primarily drawn to the textual content (i.e., titles) of items, although this may not always be the case.
That is the reason we believe TCF with only frozen text encoders performs on par with IDCF is surprising as IDCF can implicitly learn latent factors beyond textual features but feature representation pre-extracted from a frozen NLP  encoder cannot. Furthermore, we notice that SASRec with a fine-tuned text encoder can clearly outperform IDCF on all three datasets. However, as mentioned, such end-to-end training using a text encoder is computationally very expensive, despite its effectiveness.

\textbf{The answer to Q3 is that, for \textit{text-centric} recommendation, TCF with the seq2seq based SASRec backbone and utilizing a 175B parameter ``frozen'' LLM can achieve similar performance to standard IDCF, even for warm item recommendation. However, even by retraining a super-large LLM item encoder,  TCF with a DSSM\footnote{A very recent study~\citep{rajput2023recommender} suggested that standard CTR models, such as DSSM and DeepFM~\citep{guo2017deepfm}, may be replaced by the seq2seq generative architecture. This means seq2seq model may have a chance to be a mainstream recommendation architecture.
} backbone has little chance to compete with its corresponding IDCF.  The simple IDCF still remains a highly competitive approach in the warm item recommendation setting. } If the computation can be reduced, joint training of the sequential recommender backbone (i.e., SASRec) and its LLM text encoder can lead to markedly  better results than both IDCF and its frozen LLM counterpart.

\section{Q4: How close is the TCF paradigm to a universal recommender model?}
 In this paper, we are particularly interested in comparing with the dominant IDCF paradigm. This is because ID features (including userID and itemID) are considered to be a major obstacle for transferable or foundation recommender models as they cannot be easily shared between different recommender systems due to privacy and security concerns~\citep{yuan2023go,hou2022towards,rajput2023recommender,wang2022transrec,ding2021zero,shin2021one4all}. We argue that to achieve foundation models in recommender systems may require satisfying two conditions,  as illustrated in Figure~\ref{fig:foundation}: (1) abandoning userID\footnote{UserID can be represented by a sequence of interacted itemIDs by the user (e.g., in the sequential recommender model), so the key challenge is the itemID.} and itemID features, and (2) achieving effective transferability across domains and platforms. Based on the above results, we conclude that for text-centric recommender systems, TCF-based sequential recommender models can basically substitute IDCF methods. However, regarding (2), it remains unknown whether TCF has impressive transfer learning ability,  when its item representations are extracted from an extremely large LLM.

\begin{table}[t]
\centering
\caption{Zero-shot recommendation accuracy (HR@10). 175B\textsubscript{zero} means zero-shot accuracy of TCF with 175B LLM. ‘train’ is to retrain TCF on each downstream dataset.
}
\label{tab:transcompare}
\begin{tabular}{p{2.4cm}<{\centering} p{1.5cm}<{\centering} p{1.5cm}<{\centering} p{1.5cm}<{\centering}}
\toprule
\multirow{1}{*}{Model} & \multicolumn{1}{c}{MIND} & \multicolumn{1}{c}{HM} & \multicolumn{1}{c}{QB} \\
\midrule  
Random & 0.02 & 0.01 & 0.18 \\
175B\textsubscript{zero} & 0.13 & 0.39& 4.30 \\
175B\textsubscript{train} &20.24 &11.11 &29.90 \\

\bottomrule
\end{tabular}
\end{table}

Inspired by the remarkable success of zero-shot learning in NLP, our goal is to assess the more challenging zero-shot transfer learning capability of TCF, considering that items with text features may be inherently transferable. Following~\citep{ding2021zero}, we first  pre-train a SASRec-based TCF model with the 175B parameter frozen LLM as item encoder on the large-scale Bili8M dataset.
We then directly evaluate this pre-trained model in the testing set of MIND, HM and QB\footnote{QQ Browser (QB) is a feed recommendation dataset from which we extracted short video titles, similar to items from Bili. It contains 5546 items 17809 users and 137979 interactions.}. The results, presented in Table~\ref{tab:transcompare}, indicate that while TCF models outperform random item recommendation by achieving an accuracy improvement of 6-40x, they still fall notably short of TCF models that have been retrained on the new data. We note that user behaviors in the source Bili8M dataset may differ significantly from HM and MIND datasets due to their distinct contexts of e-commerce and news recommendation scenarios. However, it is similar to that of QB, as both involve similar types of item recommendations.

\textbf{The answer to Q4 is that while TCF models with LLMs do exhibit  a certain degree of transfer learning capability, they still  fall significantly  short of being a universal recommender model, as we had initially envisioned.}  For a universal recommendaton model, not only should item representations be transferable, but also the matching relationship between users and items needs to be transferable. However, the matching relationship is closely related to the exposure strategy of a specific recommender platform.

Therefore, compared to NLP and computer vision (CV), the transferability of recommender models is even more challenging. This also explains why, up until now, there haven't been any pre-trained models in the field of recommender systems that have attained the same level of prominence and recognition as BERT and ChatGPT in the NLP field.

For instance, the lack of a pre-trained recommender model in the HuggingFace library that can support various recommendation scenarios (similar or dissimilar) further reinforces this point.
However, this does not necessarily indicate that TCF have no potential  to become a universal recommender model. It will require the collective effort of the entire recommender system community. This includes utilizing highly diverse and extremely large pre-training datasets~\citep{ni2023contentdriven}, employing advanced training and transfer learning techniques, and engaging in deeper considerations for fair evaluation.

\section{Benchmarking TCF with other LLMs}
\label{section-benchmark}
OPT175B, nearly the largest open-source LLMs, developed by Meta AI, was trained using about 1000 80G A100 GPUs, positioning it as a comparable alternative to the closed-source GPT-3 model. During the course of this study, several new LLMs have emerged, including LaMDa~\citep{thoppilan2022lamda}, PaLM~\citep{chowdhery2023palm}, Gopher~\citep{rae2021scaling}, LLama~\citep{touvron2023llama, touvron2023llama2}, Mistral~\citep{jiang2024mixtral}, Falcon~\citep{almazrouei2023falcon} and MPT~\citep{MosaicML2023Introducing} etc.  

Here we are interested in benchmarking their performance as item encoders for recommender systems,  even though they have been evaluated on numerous benchmarks for NLP tasks.
Please note that we are unable to evaluate LaMDa, PaLM, and Gopher in our study as they are not open-source models.

Table~\ref{tab:LMcompare} presents the benchmark results of TCF with 10 language models, including the lightweight word2vec, medium-sized BERT, relatively large T5, and  large LM LLama2 etc. These language models represent the culmination of 10 years of research in the NLP community. Our results  reveal that TCF with OPT175B, the largest LLMs, outperforms all other models in terms of item recommendation across all three datasets. Moreover, we observed that TCF with recently proposed LLMs, including LLama2, Falcon,  Mistral, and MPT, demonstrates exceptional performance compared to smaller item encoder models such as word2vec and BERT.

\section{Discussion}
\textbf{\textit{Novelty clarification:}} our work was highly inspired by a recent study~\citep{yuan2023go}, but we have made several significant contributions.
First, ~\citep{yuan2023go} primarily focused on smaller LMs such as BERT and RoBERTa, which had parameter sizes around 100 million. In contrast, our research investigates the scaling effect of super large LMs ranging from 100 million to 175 billion parameters. Extensively studying LLMs with hundreds of billions of parameters is a non-trivial task that requires significant efforts in terms of both technical engineering and computational resources. 
To the best of our knowledge, this paper represents the pioneering effort in exploring the use of exceptionally large language models as item encoders for the top-$N$ item recommendation task. It also provides the first confirmed evidence of the positive scaling effects of LLMs for item recommendation (from implicit feedback).

Second, while both papers compare TCF with IDCF, our study arrives at a distinct  conclusion. ~\citep{yuan2023go} claimed that TCF can only compete with IDCF when the text encoders or LMs are jointly fine-tuned. However, our work discovered that when using a super large LLM, the frozen text encoder with pre-extracted offline representation can already be competitive with ID embeddings. 
This finding represents a significant progress, as fine-tuning very large LLMs  could require 100 or 1,000 times more computation, which is often impractical for large-scale real-world applications. Therefore, this is regarded as a remarkable finding that holds significant implications for recommender systems in terms of moving away from relying on explicit itemID embeddings.

Beyond these contributions, we have also studied the zero-shot transfer learning effects of TCF with super large LLMs pre-trained on a large-scale  dataset (Bili8M), albeit with surprisingly unexpected results. At last, we have benchmarked numerous well-known open-source LLMs for recommender systems, laying the foundation for future research.

The results of this research not only validate some previous findings derived from smaller LMs but also uncover new insights beyond existing studies. These experimental outcomes, regardless of their positive or negative nature, call for further contemplation and discussion within the research community.  It is worth emphasizing that these discoveries were made possible through extensive and costly empirical studies,making a significant contribution to the existing literature.

\begin{table}[t]
\centering
\caption{TCF's results (HR@10)  with renowned text encoders in the last 10 years. Text encoders are frozen and the SASRec backbone is used. Notable  advances in NLP benefit RS.}
\label{tab:LMcompare}
\begin{tabular}{p{1.8cm}<{\centering} p {1.2cm}<{\centering} p{1.2cm}<{\centering} p{1.2cm}<{\centering} p{1.2cm}<{\centering}}
\toprule
Model &Date &MIND &HM &Bili \\
\midrule
word2vec  &2013                         & 15.21 & 8.08  & 2.66 \\
\midrule
BERT\textsubscript{\textit{large}} &2018    & 18.99 & 9.68  & 3.56 \\
T5\textsubscript{\textit{XXL}}  &2019    & 19.56 & 9.21  & 4.81 \\
\midrule
LLAMA2\textsubscript{\textit{7B}} &2023 & 19.78 & 9.45 & 6.41 \\
LLAMA2\textsubscript{\textit{13B}} &2023 & 19.68 & 9.70 & 6.63 \\
LLAMA2\textsubscript{\textit{70B}} &2023 & 19.84 & 9.80 & 6.79 \\
MPT\textsubscript{\textit{30B}} &2023 & 19.64 & 9.71 & 6.26 \\
FALCON\textsubscript{\textit{40B}} &2023 & 19.92  & 10.13 & 5.64 \\
MISTRAL\textsubscript{\textit{7B}} &2023 & 19.79 & 10.67 & 5.56 \\
MISTRAL\textsubscript{\textit{8*7B}} &2023 & 19.74 & 10.71 & 5.98 \\
\midrule
\textbf{OPT\textsubscript{\textit{175B}}} &2022   & 20.24 & 11.11 & 7.05 \\ 
\bottomrule
\end{tabular}
\vskip -0.1in
\end{table}

\textbf{\textit{Limitations:}}
Although this paper presents a solid empirical study, it has a potential limitation: all observations are derived from offline data, whose evaluation may be influenced by exposure bias~\citep{liumeasuring,khenissi2020modeling}. The applicability of these findings in the practical online environment remains further verification, as researchers in academia often face limitations in accessing online data. In addition, online recommender systems are complex compositions that incorporate multiple algorithms, stages, and strategies, and their evaluations are academically unreproducible. 
The pursuit of methods for conducting fair and reproducible evaluations continues to be a very challenging problem in the RS community~\citep{gao2022fair,li2023fairness}. We hope that our findings  will stimulate further contemplation among researchers and help address the limitations in future studies.

\section{Conclusion}
This paper does not describe a new text recommender algorithm. Instead, it extensively explores the performance limits and several core issues of the prevailing text-based collaborative filtering (TCF) techniques by  scaling its LLMs item encoder. From a positive perspective, TCF still holds untapped potential and has room for further improvement as the representation capacity of larger NLP models advances. However, on the other hand, even with item encoders consisting of tens of billions of parameters, re-adaptation to new data remains necessary for optimal recommendations. Furthermore, the current state-of-the-art TCF models do not exhibit the anticipated strong transferability, suggesting that building large foundation recommender models may be more challenging than in the field of NLP. Nonetheless, TCF with text encoders of 175 billion parameters is already a significant leap  forward, as it fundamentally challenges decade-long dominance of the ID-based CF paradigm, which is considered the biggest obstacle to  developing universal ``one-for-all" recommender models, although not the only one.

\newpage
\appendix

\section{Hyper-parameter tuning}
\label{Hyper-parameter}
Before tuning hyper-parameters for TCF, we grid search IDCF on each dataset as a reference.
Specifically, we search for learning rates within the range of \{\textit{1e-3, 1e-4, 1e-5, 5e-5}\}  and hidden dimensions from  \{\textit{64, 128, 256, 512, 1024, 2048}\}  for both DSSM and SASRec; we search batch size within \{\textit{64, 128, 256, 512}\} for SASRec and \{\textit{1024, 2048, 4096}\} for DSSM; 
we set a fixed dropout rate of 0.1, and tune the weight decay within \{\textit{0.01, 0.1}\};
we search the number of Transformer layers in SASRec within  \{\textit{1, 2, 3, 4}\}, and the number of attention heads within \{\textit{2, 4, 8}\}.  After determining the optimal hyper-parameters for IDCF, we search the TCF around these optimal values with the frozen text encoder (using the 125M variant) by the same stride.
To ensure a fair comparison of the scaling effect, we employ the same hyper-parameters for all TCF models with different sizes of frozen text encoder (i.e., pre-extracted features).
For TCF models with expensive E2E learning of text encoders, we kept the optimal hyper-parameters the same as those with frozen encoder, except for the learning rates. 
We separately tune the learning rate, as larger text encoders typically require a smaller learning rate.  The details are given below.
AdamW~\cite{loshchilov2017decoupled} is applied for all models.


\begin{table}[h]
\centering
\caption{Optimal hyper-parameters for IDCF, including learning rate ($lr$), embedding size  ($k$), batch size ($bs$), the number of Transformer layers ($l$), the number of attention heads ($h$), and weight decay is 0.1 for all. The  dimension of feed forward layer in Transformer block is   $4 \times k$.
}
\vskip -0.05in
\label{IDHyper}
\begin{tabular}{p{0.8cm}<{\centering} p{0.6cm}<{\centering} p{0.3cm}<{\centering} p{0.3cm}<{\centering} p{0.2cm}<{\centering} p{0.2cm}<{\centering}  p{0.6cm}<{\centering} p{0.6cm}<{\centering} p{0.3cm}<{\centering} p{0.2cm}<{\centering} p{0.2cm}<{\centering}}
\toprule
\multirow{2}{*}{\textbf{Data}} &\multicolumn{5}{c}{SASRec} &\multicolumn{5}{c}{DSSM}\\
\cmidrule(lr){2-6} \cmidrule(lr){7-11}
&\multicolumn{1}{c}{$lr$} &\multicolumn{1}{c}{$k$} &\multicolumn{1}{c}{$bs$} &\multicolumn{1}{c}{$l$} &\multicolumn{1}{c}{$h$}  &\multicolumn{1}{c}{$lr$} &\multicolumn{1}{c}{$k$} &\multicolumn{1}{c}{$bs$} &\multicolumn{1}{c}{$l$} &\multicolumn{1}{c}{$h$}\\
\midrule
MIND &1e-4 &512 &64  &2 &2 &1e-5 &256  &4096 &2 &2 \\
HM   &1e-3 &128 &128 &2 &2 &1e-4 &1024 &1024 &2 &2 \\
Bili &1e-3 &128 &256 &2 &2 &1e-3 &1024 &1024 &2 &2  \\
\bottomrule
\end{tabular}
\end{table}


\begin{table}[h]
\centering
\caption{Optimal hyper-parameters for  TCF with frozen text encoder, weight decay is 0.1 for all. }
\vskip -0.05in
\label{TCFHyper}
\begin{tabular}{p{0.7cm}<{\centering} p{0.6cm}<{\centering} p{0.3cm}<{\centering} p{0.3cm}<{\centering} p{0.2cm}<{\centering} p{0.2cm}<{\centering}  p{0.6cm}<{\centering} p{0.6cm}<{\centering} p{0.3cm}<{\centering} p{0.2cm}<{\centering} p{0.2cm}<{\centering}}
\toprule
\multirow{2}{*}{\textbf{Data}} &\multicolumn{5}{c}{SASRec} &\multicolumn{5}{c}{DSSM}\\
\cmidrule(lr){2-6} \cmidrule(lr){7-11}
&\multicolumn{1}{c}{$lr$} &\multicolumn{1}{c}{$k$} &\multicolumn{1}{c}{$bs$} &\multicolumn{1}{c}{$l$} &\multicolumn{1}{c}{$h$}  &\multicolumn{1}{c}{$lr$} &\multicolumn{1}{c}{$k$} &\multicolumn{1}{c}{bs} &\multicolumn{1}{c}{$l$} &\multicolumn{1}{c}{$h$}\\
\midrule
MIND &1e-4 &512 &64 &2 &2  &1e-5 &256  &4096 &2 &2 \\
HM   &1e-4 &512 &64 &2 &2  &1e-3 &1024 &1024 &2 &2 \\
Bili &1e-3 &128 &64 &2 &2  &1e-3 &512  &1024 &2 &2 \\
\bottomrule
\end{tabular}
\end{table}

\vspace{-0.1in}

\begin{table}[h]
\centering
\caption{The learning rate of item encoder for  TCF with E2E learning. The search range is suggested by the original paper of OPT.}
\vskip -0.1in
\label{Ft_TCF_LR}
\begin{tabular}{p{0.5cm}<{\centering} p{0.6cm}<{\centering} p{0.6cm}<{\centering} p{0.55cm}<{\centering} p{0.55cm}<{\centering} p{0.6cm}<{\centering} p{0.6cm}<{\centering} p{0.55cm}<{\centering} p{0.55cm}<{\centering}}
\toprule
\multirow{2}{*}{\textbf{Data}} &\multicolumn{4}{c}{SASRec} &\multicolumn{4}{c}{DSSM}\\
\cmidrule(lr){2-5} \cmidrule(lr){6-9}
&\multicolumn{1}{c}{125M} &\multicolumn{1}{c}{1.3B} &\multicolumn{1}{c}{13B} &\multicolumn{1}{c}{66B}  &\multicolumn{1}{c}{125M}   &\multicolumn{1}{c}{1.3B}  &\multicolumn{1}{c}{13B} &\multicolumn{1}{c}{66B} \\
\midrule
MIND &1e-4 &1e-4 &8e-5 &3e-5 &1e-4 &1e-4 &1e-4 &1e-4 \\
HM   &1e-4 &1e-4 &1e-4 &8e-5 &1e-4 &1e-4 &1e-4 &1e-4 \\
Bili &1e-4 &1e-4 &3e-5 &3e-5 &1e-4 &1e-4 &1e-4 &1e-4 \\
\bottomrule
\end{tabular}
\end{table}

\section{Text encoder details} 
We provide the sources of the models used in our study in Table~\ref{tab:text encoders}.

\begin{table}[h]
\centering
\caption{LM-based text encoder details. Para. represent the number of parameters}
\label{tab:text encoders}
\scalebox{0.85}{
\begin{tabular}{p{1.0cm}<{\centering} p{0.8cm}<{\centering} l}
\toprule
\textbf{Name}   & \textbf{Para.} & \textbf{Source} \\
\midrule
\multirow{1}{*}{BERT}      & 340M & \url{https://huggingface.co/bert-large-uncased} \\ 
\midrule
\multirow{1}{*}{T5Encoder} & 5.5B & \url{https://huggingface.co/t5-11B} \\ 
\midrule
\multirow{9}{*}{OPT}       & 125M & \url{https://huggingface.co/facebook/opt-125m} \\ 
                           & 350M & \url{https://huggingface.co/facebook/opt-350m} \\ 
                           & 1.3B & \url{https://huggingface.co/facebook/opt-1.3b} \\  
                           & 2.7B & \url{https://huggingface.co/facebook/opt-2.7b} \\ 
                           & 6.7B & \url{https://huggingface.co/facebook/opt-6.7b} \\ 
                           & 13B  & \url{https://huggingface.co/facebook/opt-13b} \\  
                           & 30B  & \url{https://huggingface.co/facebook/opt-30b} \\  
                           & 66B  & \url{https://huggingface.co/facebook/opt-66b} \\  
                           & \multirow{2}{*}{175B}  & \parbox[t]{7cm}{\url{https://github.com/facebookresearch/metaseq/tree/main/projects/OPT}} \\     
\midrule
\multirow{3}{*}{LLAMA2}    & 7B  & \url{https://huggingface.co/meta-llama/Llama-2-7b-hf} \\ 
                           & 13B & \url{https://huggingface.co/meta-llama/Llama-2-13b-hf} \\ 
                           & 70B & \url{https://huggingface.co/meta-llama/Llama-2-70b-hf} \\  
\midrule
\multirow{1}{*}{MPT}       & 30B & \url{https://huggingface.co/mosaicml/mpt-7b} \\ 
\midrule
\multirow{1}{*}{FALCON}    & 40B & \url{https://huggingface.co/tiiuae/falcon-40b} \\ 
\midrule
\multirow{2}{*}{MISTRAL}   & 7B  & \url{https://huggingface.co/mistralai/Mistral-7B-v0.1} \\ 
                           & 8*7B & \url{https://huggingface.co/mistralai/Mixtral-8x7B-v0.1} \\ 
\bottomrule
\end{tabular}
}
\end{table}

\vspace{0.3em}

\section{TCF results on Bili8M} 

\begin{figure}[h]
    \centering
    \includegraphics[width=0.45\textwidth]{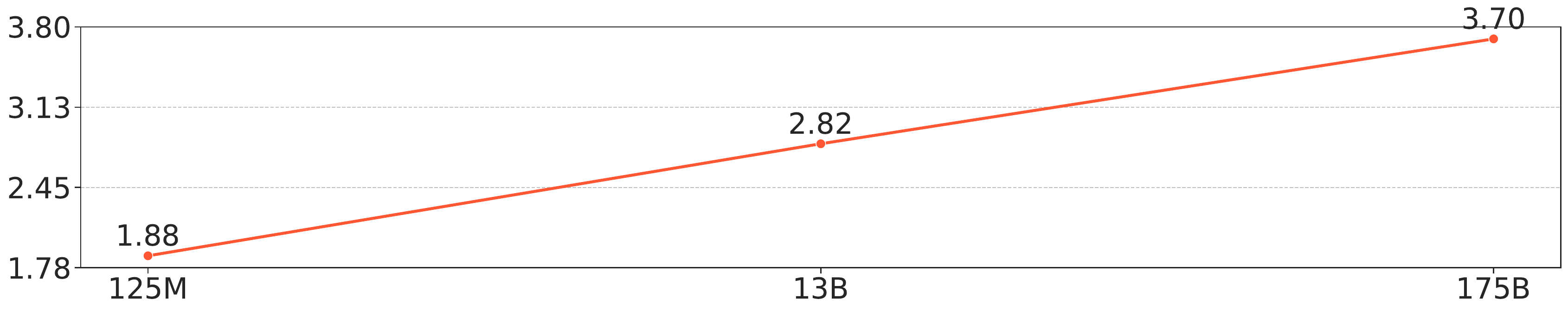} 
    \caption{TCF’s performance (y-axis: HR@10(\%)) of 3 item encoders with increased sizes (x-axis) on Bili8M. SASRec is used as the backbone. LLM is frozen.} 
    \label{fig:TCFlargeBili8M} 
\end{figure}

\clearpage

\bibliographystyle{ACM-Reference-Format}
\balance
\bibliography{sample-base}

\section*{GenAI Usage Disclosure}
The authors affirm that no generative AI systems were used for data collection, manuscript preparation, figure generation, or experimental analysis in this work. We note, however, that ChatGPT was involved solely as the subject of investigation within our experiments. Its role was limited to being evaluated and analyzed as part of the study, rather than being employed as a tool for writing or content generation in the manuscript.

\end{document}